# Free Internal Waves in Polytropic Atmospheres

Mikhail I. Ivanov[*]

## ABSTRACT

Free internal waves in polytropic atmospheres are studied (polytropic atmosphere is such one that the temperature of gas linearly depends on altitude). We suppose gas to be ideal and incompressible. Also, we regard the atmosphere of constant height with the "rigid lid" condition on its top to filter internal waves. If temperature, density and pressure of such undisturbed atmosphere do not depend on latitude and longitude then the internal waves are harmonic with apriori unknown eigenfrequencies, the problem permits separation of variables and reduces to the system of two ODE's. The first ODE (the Laplace's tidal equation) is analyzed by author earlier. The second ODE determines the vertical structure of the waves to be considered and has analytical solution for polytropic atmospheres. There are 6 dimensionless numbers, 2 for the Laplace's tidal equation and 4 for the vertical structure equation. The solution is a countable set of the eigenfrequencies and eigenfunctions of the vertical structure equation; every eigenfrequency/eigenfunction corresponds to its own countable set of the eigenfrequencies and eigenfunctions of the Laplace's tidal equation.

Parametric analysis of the problem has been done. It shows that there exists the solution weakly depending on altitude-temperature variations and the atmosphere's height for parameters modelling the Earth's troposphere (with the "rigid lid" between the troposphere and the tropopause). The natural periods of internal waves have been obtained for this case.

## INTRODUCTION

It is well known that the solar semidiurnal tide has essentially greater amplitude in the Earth's atmosphere than the lunar one (despite of the lunar tidal force is 2.2 times more the solar one). Furthermore, that amplitude is 80 times more its static estimated value [1, 2]. It may be explained by a resonant mechanism, i.e., by the fact that the Earth's atmosphere has a natural oscillation with a period near to 12 hours. In this connection investigation of the natural oscillations of the Earth's atmosphere is of special interest. In addition, the natural oscillations of the Earth's atmosphere have been observed after vast atmospheric disturbances as Krakatau explosion [3].

This problem has been reduced to integration of two related equations – the Laplace's tidal equation (LTE) and the vertical structure equation (VSE). There exists voluminous literature on LTE, for example [4-6]. LTE depend on physical data through two numbers, one is setting and one is searching. VSE has another character. The equation form depends on an altitude-temperature function (a temperature as a function of altitude) that is rather complex but may be approximated by several simple pieces. This approach is the basis for integration of the problem of the natural and forced oscillations of the Earth's atmosphere [7]. However, using of various altitude-temperature functions leads us to essentially different results. Reason of that is unsteadiness of some modes about altitude-temperature function small variations [1]. Those modes have been generated by problem formulation specialities. Obviously, they are fictitious and do not exist in the Earth's atmosphere. The problem is complicated by the fact that rather comprehensive studies had been done only for an infinite isothermal atmosphere [3].

---

[*] Russian Academy of Sciences, A. Ishlinsky Institute for Problems in Mechanics, Moscow, Russia;
*e-mail*: m-i-ivanov@mail.ru

Therefore, to determine a general character of interesting wave motions it is worthwhile to restrict ourselves to atmospheric models with line altitude-temperature functions (so-called polytropic atmospheres) and bring it under close study. Note that an isothermal atmosphere is a limit case of a polytropic one. Usually it has examined an atmosphere without top (an infinite atmosphere) in the literature [3, 7]. This is scarcely valid since the Earth's atmosphere changes its gas composition at altitude of 100 km in respect of that in lower regions [8] and, apparently, dynamo effects occasioned by ionization are of importance in higher regions [2]. Whereas, we take an interest in internal waves and we use the "rigid lid" condition at the upper boundary of the atmosphere. It means that no vertical motions take place between atmospheric layers that is peculiar to a demarcation line between stable and unstable stratified atmospheric regions (for example, the troposphere and tropopause [8]).

## FORMULATION OF THE PROBLEM

We consider a compressible atmosphere of ideal gas rotating with a planet with an angular velocity $\omega$. The temperature $T$ of that gas depends on altitude $z$ linearly ($\frac{dT}{dz} = \alpha T(0) = \text{const}$, $T(0)$ is the planet's surface temperature). Such atmosphere is called polytropic. We suppose the height of the atmosphere is small in respect of the planet's radius and the atmospheric composition does not change. We suppose the planet's surface is flat and neglect dissipative processes. An undisturbed atmosphere is in an equilibrium state while a pressure, density and temperature does not depend on latitude and longitude:

$$p = p(0)\exp\left(-\frac{\mu_a g}{R}\right)\int_0^z \frac{dz}{T(z)}$$

$$\rho = \frac{p}{RT} \tag{1}$$

Here, $p(0)$ is the undisturbed pressure on the planet's surface, $\mu_a$ is the average molecular mass of an atmospheric gas, $R$ is the gas constant, $g$ is the gravitational acceleration to be accepted constant because of the atmosphere's height is small.

We study small disturbances of the state (1) to be considered linearized relative to (1). We accept even a disturbed atmosphere is in a hydrostatic and local thermodynamic equilibrium state. The basic system includes three linearized equations of motion, the continuity equation, the equation of a heat influent by adiabatic compression and the linearized state equation of ideal gas. In the spherical coordinate system it is written:

$$\frac{\partial u}{\partial t} - 2\omega v \cos\theta = -\frac{1}{a}\frac{\partial}{\partial\theta}\left(\frac{\delta p}{\rho}\right)$$

$$\frac{\partial v}{\partial t} + 2\omega u \cos\theta = -\frac{1}{a\sin\theta}\frac{\partial}{\partial\varphi}\left(\frac{\delta p}{\rho}\right)$$

$$\frac{\partial \delta p}{\partial z} = -g\delta\rho$$

$$\frac{D\rho}{Dt} = -\rho\left(\frac{1}{a\sin\theta}\frac{\partial(u\sin\theta)}{\partial\theta} + \frac{1}{a\sin\theta}\frac{\partial v}{\partial\varphi} + \frac{\partial w}{\partial z}\right)$$

$$\frac{1}{\gamma-1}\frac{DT}{Dt} = \frac{T}{\rho}\frac{D\rho}{Dt}$$

$$\frac{\delta p}{p} = \frac{\delta T}{T} + \frac{\delta\rho}{\rho} \qquad (2)$$

Here, $\theta$ is colatitude, $\varphi$ is longitude, $\delta T$, $\delta p$, $\delta\rho$ is the temperature, pressure and density disturbances, correspondingly, and $a$ is the planet's radius.

In such atmosphere there exist free internal waves with pressure evolution of the form $\frac{Dp}{Dt} = -\gamma p(z) y(z) \Theta(\theta) \exp(i(n\varphi + \sigma t))$, where $\gamma$ is the adiabatic number, $n$ is the latitudinal wavenumber, and $\sigma$ is the internal wave frequency [1, 2]. Those waves are described by the system:

$$f\frac{d}{d\mu}\left(\frac{1-\mu^2}{f^2-\mu^2}\frac{d\Theta}{d\mu}\right) + \left(\frac{n(f^2+\mu^2)}{(f^2-\mu^2)^2} - \frac{n^2 f}{(1-\mu^2)(f^2-\mu^2)} + \beta f\right)\Theta = 0$$

$$\frac{d^2 y}{dx^2} + \left(\Omega_0\beta(\kappa+\lambda)\exp(\lambda x) - \frac{1}{4}\right)y = 0 \qquad (3)$$

Here, $\mu = \cos\theta$, $x = \frac{1}{\lambda}\ln(1+\alpha z)$, $f = \frac{\sigma}{2\omega}$, $\lambda = \frac{\alpha RT(0)}{\mu_a g}$, $\Omega_0 = \frac{RT(0)}{4\mu_a \omega^2 a^2}$, $\kappa = \frac{\gamma-1}{\gamma}$. Note that the function $\Theta$ is dimensionless but the function $y$ has dimensionality 1/sec.

The system (3) is to be supplemented by the boundary conditions:

$$|\Theta(\pm 1)| < \infty$$

$$\Lambda y(0) = 0$$

$$\Lambda y\left(\frac{1}{\lambda}\ln\frac{T(Z)}{T(0)}\right) = 0 \qquad (4)$$

Here, $\Lambda = \frac{d}{dx} + \left(\Omega_0\beta\exp(\lambda x) - \frac{1}{2}\right)$, $Z$ is the height of the atmosphere.

The disturbances are in the form:

$$\delta p = \frac{\gamma p(0)}{i\sigma\Omega_0\beta}\exp\left(-\lambda x - \frac{x}{2}\right)\left(\frac{d}{dx}-\frac{1}{2}\right)y\,\Theta(\mu)\exp(i(\sigma t + n\varphi))$$

$$\delta\rho = \frac{\gamma\mu_a p(0)}{i\sigma\Omega_0\beta RT(0)}\exp\left(-2\lambda x - \frac{x}{2}\right)\times$$

$$\times\left((1+\lambda)\frac{d}{dx}+(\kappa+\lambda)\Omega_0\beta\exp(\lambda x)-\frac{1+\lambda}{2}\right)y\,\Theta(\mu)\exp(i(\sigma t + n\varphi))$$

$$\delta T = -\frac{\gamma T(0)}{i\sigma\mu_a\Omega_0\beta}\exp(x/2)\left(\lambda\frac{d}{dx}+(\kappa+\lambda)\Omega_0\beta\exp(\lambda x)-\frac{\lambda}{2}\right)y\,\Theta(\mu)\exp(i(\sigma t + n\varphi))$$

$$u = -\frac{\gamma a}{\beta}\exp(x/2)\left(\frac{d}{dx}-\frac{1}{2}\right)y\frac{1}{(f^2-\mu^2)\sqrt{1-\mu^2}}\left((1-\mu^2)\frac{d}{d\mu}-n\frac{\mu}{f}\right)\Theta(\mu)\exp(i(\sigma t + n\varphi))$$

$$v = -\frac{i\gamma a}{\beta}\exp(x/2)\left(\frac{d}{dx}-\frac{1}{2}\right)y\frac{1}{(f^2-\mu^2)\sqrt{1-\mu^2}}\left(\mu(1-\mu^2)\frac{d}{d\mu}-n\right)\Theta(\mu)\exp(i(\sigma t + n\varphi))$$

$$w = \frac{\gamma RT(0)}{g\mu_a\Omega_0\beta}\exp(x/2)\left(\frac{d}{dx}+\Omega_0\beta\exp(\lambda x)-\frac{1}{2}\right)y\,\Theta(\mu)\exp(i(\sigma t + n\varphi)) \quad (5)$$

We can see that the pressure and density disturbances grow up in the increase of altitude but the temperature and velocity disturbances fall down.

## EXACT SOLUTIONS

The first equation of (3) is LTE. Its numerical integration technique for different $\beta$ (including negative ones) had been developed in [4-5]. The second equation is VSE. It has the exact solution that depends on signs of $\lambda$ and $\Omega_0\beta(\kappa+\lambda)$. In according to a definition $\Omega_0$ is always positive.

For $\lambda > 0$, $\beta > 0$ we have [9]:

$$y = C_1 J_{1/\lambda}(F(x)) + C_2 Y_{1/\lambda}(F(x)) \quad (6)$$

Here, $F(x) = \frac{2\sqrt{\Omega_0\beta(\kappa+\lambda)}}{\lambda}\exp\left(\frac{\lambda x}{2}\right)$, $C_i$ is some real constants defining from (4). For $\beta < 0$ there does not exist solutions to satisfy the boundary conditions therefore we do not write corresponding eigenfunctions.

Formally, for $-\kappa < \lambda < 0$, $\beta > 0$ there has the same solution as for $\lambda > 0$, $\beta > 0$, but the cylindrical functions have complex values yet. Whereas, obviously, solutions of VSE are always real, therefore, it is necessary that $C_i$ are complex in such a way that the solution (6) would be real. Hence, we obtain some relation between real and imaginary parts of $C_i$, and then the solution (6) have been written:

$$y = \left(C_1\cos\frac{\pi}{-\lambda}+C_2\sin\frac{\pi}{-\lambda}\right)J_{1/\lambda}(-F(x)) +$$
$$+\left(-C_1\sin\frac{\pi}{-\lambda}+C_2\cos\frac{\pi}{-\lambda}\right)Y_{1/\lambda}(-F(x)) \quad (7)$$

Here, the constants are re-symbolized in such a way that new $C_i$ would be real.

As before, we do not write eigenfunctions for $\beta < 0$ too since there does not exist solutions to satisfy the boundary conditions.

For $\lambda < -\kappa$ we has the same solution as for the previous case ($-\kappa < \lambda < 0$) with the substitution of $-\beta$ for $\beta$.

The parameters defining specialities of LTE solutions (i.e., natural oscillation horizontal structure) are the dimensionless numbers $\beta$, $f$ and latitudinal wavenumber $n$. The parameters defining specialties of VSE solutions (i.e., natural oscillation vertical structure) are the dimensionless numbers $\lambda$, $\kappa + \lambda$, $\Omega_0 \beta$ and $\frac{1}{\lambda} \ln \frac{T(z)}{T(0)}$. As one can see, LTE and VSE are connected with each other through the numbers $\beta$ and $\Omega_0 \beta$ (remind that $\Omega_0$ is constant for any fixed atmosphere). There is the countable set $\Omega_0 \beta_k$ of VSE's eigenvalues and any Lamb number $\beta_k$ for fixed $n$ relate with the certain countable set of the dimensionless frequencies $f_l$.

A limit case of a polytropic atmosphere is an isothermal one ($\alpha = 0$). In that case the solutions of VSE is trigonometric or exponential functions. The spectrum is:

$$\Omega_0 \beta_0 = 1 - \kappa$$

$$y = \exp\left(-\left(\frac{1}{2} - \kappa\right)x\right) \qquad (8)$$

$$\Omega_0 \beta_k = \frac{1}{\kappa}\left(\left(\frac{\pi k}{X}\right)^2 + \frac{1}{4}\right), \ X = \frac{\mu_a g}{RT(0)} Z, \ k \geq 1$$

$$y = -4\pi k \kappa X \cos \frac{\pi k x}{X} + \left(4\pi^2 k^2 + (1 - 2\kappa) X^2\right)\sin \frac{\pi k x}{X} \qquad (9)$$

It includes the mode (8), monotonically decreasing with altitude (emphasize that the eigenfrequency does not depend on the height), and the countable set of the eigenfunctions (9) with alternating signs.

In the real planet's atmospheres $\lambda$ is small (for example, $\lambda = -0.19$ even in the Earth's troposphere where the temperature strongly decreases with the increase of altitude) and the atmosphere's height is fairly small. Therefore, from the geophysical point of view the most important case is of $\lambda \to 0$. One can show that the cylindrical functions of the solutions (6) and (7) tend to trigonometric and hyperbolic ones. Finally, we have the spectrum of the isothermal case (8)-(9). At this point it is important that the function $\kappa \Omega_0 \beta \exp(\lambda x) - \frac{1}{4}$ would keep the sign on the entire interval of altitudes $x$. It is just a condition of smallness of the atmosphere's height.

## RESULTS

In the case of $\alpha \neq 0$ VSE's integration reduces to solving of the certain transcendental equation to have been obtained by the substitution of the solutions (6) and (7) in the boundary conditions (4). Numerical investigation of the solutions of that transcendental equation shows that for $\lambda > -\kappa$ the spectrum is indistinguishable from the isothermal case in a qualitative sense and for $\lambda < -\kappa$ the mode (8) vanishes but the modes (9) with alternating signs remain. The results are in the table. Typical eigenfunctions (for $\lambda > -\kappa$) have been shown at the figure. For the case of $\lambda < -\kappa$ we have the same figure exact the mode I.

| λ | -0.5 | -0.35 | -0.1 | -0.01 | 0 | 0.01 | 0.1 | 0.5 |
|---|---|---|---|---|---|---|---|---|
| $X=1$ | - | - | 0.747 | 0.718 | 0.714 | 0.711 | 0.681 | 0.548 |
|  | -60.82 | -187.3 | 57.17 | 36.88 | 35.42 | 34.06 | 25.04 | 10.36 |
|  | -237.4 | -734.6 | 224.7 | 144.8 | 139.0 | 133.7 | 98.05 | 39.55 |
|  | -531.5 | -1647 | 504.0 | 324.7 | 311.8 | 299.7 | 219.7 | 88.21 |
| $X=5$ | - | - | 0.822 | 0.726 | 0.714 | 0.702 | 0.581 | 0.111 |
|  | -8.828 | -21.91 | 4.363 | 2.391 | 2.257 | 2.133 | 1.366 | 0.394 |
|  | -27.04 | -64.60 | 12.47 | 6.793 | 6.402 | 6.041 | 3.750 | 0.782 |
|  | -55.83 | -134.0 | 26.04 | 14.13 | 13.31 | 12.55 | 7.717 | 1.394 |
| $X=25$ |  |  | 0.837 | 0.730 | 0.714 | 0.696 | 0.098 |  |
|  |  |  | 2.937 | 1.095 | 0.930 | 0.933 | 0.170 |  |
|  |  |  | 4.683 | 1.295 | 1.096 | 1.173 | 0.249 |  |
|  |  |  | 6.648 | 1.611 | 1.372 | 1.504 | 0.337 |  |

*Table. Eigenfrequencies $\Omega_0 \beta_k$ for different temperature gradients and atmosphere's heights, $\kappa = 2/7$. Note. We do not calculate non-physical cases of combination of great atmosphere's heights and great temperature gradients and we retain empty columns at that table places.*

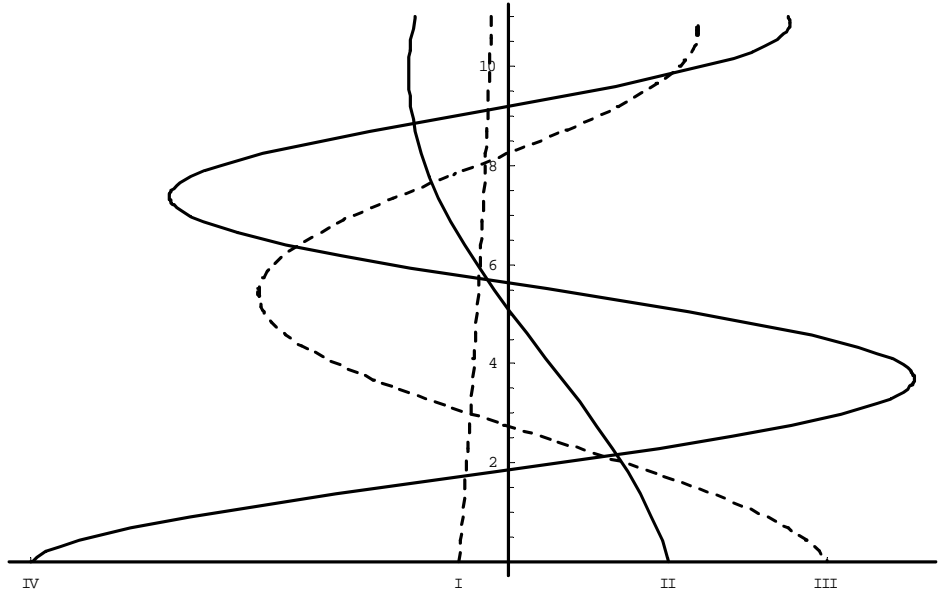

*Figure. Pressure disturbance amplitudes $\delta p$ as a function of altitude for $\lambda = -0.19$ and $X = 1.5$, that correspond to a polytropic atmosphere with a height 11 km and a temperature gradient –6.5 K/km (the Earth's troposphere modelling). Eigenvalues $\Omega_0 \beta_k$ is: I – 0.803, II – 55.53, III – 213.8, IV – 477.6.*

In general, we conclude that one can use an isothermal model for small altitudinal temperature gradients.

The null eigenvalue $\Omega_0 \beta_0$ for small $\lambda$ depend on $Z$ weakly. Whereas, frequencies of the modes with alternating signs for any $\lambda$ essentially grow up with an increase of the height. Hence, one can conclude that the null eigenvalue oscillation is unique one to have been steady about altitude-temperature function small variations and, apparently, to take place in the real atmospheres. It has been corresponded to a decreasing function $y(x)$ of constant sign.

The critical value of $\lambda$ is $-\kappa$. In that limit case there exists only one mode depending on the atmosphere's height:

$$\Omega_0 \beta_0 = \frac{\exp((1-\kappa)X)}{\exp(X)-1}$$

$$y = (\exp(\kappa X)-1)\exp\left(\frac{x}{2}\right) + (\exp(X)-\exp(\kappa X))\exp\left(-\frac{x}{2}\right) \qquad (10)$$

The eigenfunction (10) is of constant sign, but non obligatory monotonic. For fairly large $X$ it has a minimum.

With away $\lambda$ from $-\kappa$ moduli of internal wave eigenfrequencies $\Omega_0\beta_k$ decrease (weakly-marked violations of that effect take place for great $Z$ and small $\lambda$). The augmentation of the atmosphere's height $Z$ leads to the same effect (exclusive of the null mode for negative $\lambda$). Besides that, eigenfrequencies for $\lambda > -\kappa$ are positive and ones for $\lambda < -\kappa$ are negative. With the results of [5] for oscillatory energy distribution in latitudinal direction as a function of $\beta$, one can conclude that for $\lambda \to -\kappa+0$ oscillatory energy concentrates near the equator but for $\lambda \to -\kappa-0$ it does near the poles. With away from the critical value $-\kappa$ internal wave energy, in general, distributes amongst all the latitudes. The augmentation of the atmosphere's height $Z$ leads to the same effect.

## REFERENCES

1. S. Chapman and R.S. Lindzen, *Atmospheric Tides. Thermal and Gravitational*, 1970 (Dordrecht: D. Reidel).
2. S.-I. Akasofu and S. Chapman, *Solar Terrestrial Physics*, 1972 (Oxford: Clarendon Press).
3. L.A. Dikii, The Earth's Atmosphere as an Oscillatory System. *Izv. Ross. Akad. Nauk, Fizika Atmosfery i Okeana* 1, (5), 469-489 (1965) [in Russian].
4. M.I. Ivanov, Nonaxisymmetric Solution of Laplace's Tidal Equation and Rossby Waves. *Fluid Dynamics* **42**, (4), 644-653 (2007).
5. M.I. Ivanov, Horizontal Structure of Atmospheric Tidal Oscillations. *Fluid Dynamics* **43**, (3), 447-460 (2008).
6. M.S. Longuet-Higgins, The Eigenfunctions of Laplace's Tidal Equations over a Sphere. *Phil. Trans. Roy. Soc. London, Ser. A* **262**, 511-607 (1968).
7. S. Kato, Diurnal Atmospheric Oscillation. Pt. II. Thermal Excitation in the Upper Atmosphere. *J. Geoph. Res.* **71**, 3211-3214 (1966).
8. S.P. Khromov and M.A. Petrosyantz, *Meteorology and Climatology*, 2001 (Moscow: MSU) [in Russian].
9. A.D. Polyanin and V.F. Zaitsev, *Handbook of Exact Solutions for Ordinary Differential Equations*, 2003 (New York: CRC Press, Boca Raton).